\documentclass[twocolumn,letterpaper]{revtex4}

\usepackage{appendix}
\usepackage{graphicx}
\usepackage{dcolumn}
\usepackage{bm}
\usepackage{amssymb,amsmath}
\usepackage{color}
\usepackage{natbib}

\begin{document}

\title{Criteria for reliable entanglement quantification
with finite data}
\author{J. O. S. Yin and S. J. van Enk}
\affiliation{Physics Department and Oregon Center for Optics\\
University of Oregon\\
Eugene, OR 97403}
\date{\today}
\begin{abstract}
We propose one and a half criteria for determining how many measurements are needed to quantify entanglement reliably.
We base these criteria on Bayesian analysis of measurement results, and apply our methods to four-qubit entanglement, but generalizations to more qubits are straightforward.
\end{abstract}
\maketitle

\section{INTRODUCTION}\label{sect_introduction}

The study of quantum entanglement never ceases to intrigue researchers \citep{HHHH2009}, and its verification has attracted just as much attention in the quantum information community. Almost all entanglement verification methods \citep{GT2009} are designed for the situation where infinitely many data are (implicitly) assumed  to exist. The finite-data regime has not been given much attention until recently \citep{B-KYvE2010}. In that paper the main question concerned the binary decision about whether one's quantum systems are entangled or not. In the present paper we consider the task of quantifying entanglement with finite data. One of the questions we consider here is: how many measurements are needed to quantify entanglement reliably? Obviously, such a question cannot be answered in its full generality, as it will depend on what measurements are performed, on the number of qubits, and, possibly, on how accurate an estimate one wishes to have. Nevertheless, we will develop  general criteria for determining a ``sufficient'' number of measurements based on a Bayesian analysis of measurement data, which can be applied to any sorts of measurements and to any number of qubits.  Our criteria do not actually need an accuracy to be specified in advance.

The other goal of this paper is to develop Bayesian estimation methods for entanglement in nontrivial cases. In particular, we choose to simulate experiments on (mixed, entangled) four-qubit states. Ref.~\citep{B-K2010} discusses the virtues of Bayesian methods for quantum state estimation, especially as compared to maximum likelihood estimation (MLE), and here we consider that same comparison in the context of entanglement estimation.

An advantage of Bayesian methods is that error bars on entanglement measures are generated automatically. MLE can generate error bars by using a bootstrap method, where $\rho_{{\rm MLE}}$ is used to numerically generate more data, but this does not work when the number of data is small.
In Sec.~\ref{sec_differentiating} we compare these two methods of generating entanglement estimates and their error bars.
The Bayesian methods do require one to choose {\em prior} probability distributions over states. In Sec.~\ref{sec_prior} we explicitly provide two inherently different standard prior distributions in our systems both of which are numerically feasible and both of which can be applied to any number of qubits.
In Sec.~\ref{sec_multipartite} a convenient entanglement measure is introduced, that can be computed directly from the multipartite density matrices, and which can, likewise, be generalized to any number of qubits. We also briefly discuss the disadvantages of this particular measure (no known multi-partite entanglement measure is without flaws, although a very recent preprint does improve upon the situation \citep{JMG2010}. In Sec.~\ref{sec_sic-povm} we derive the relations needed for tomographic state reconstruction that are associated with a special kind of tomographically complete measurements, and in Sec.~\ref{sec_methods}  we discuss our implementation of the Metropolis-Hasting algorithm, which allows one to sample from the posterior distribution efficiently. Finally, in Sec.~\ref{sec_measurements} we give our main results and attempt to answer the questions laid out in this Introduction.

\section{Some comparisons between Maximum Likelihood Estimation and BAYESIAN METHODS}\label{sec_differentiating}

In entanglement verification experiments where tomography is adopted, maximum likelihood estimation  is widely accepted as the state estimation method of choice. Here the state that best fits the data, $\rho_{\rm MLE}$, is accepted as the best estimate of the quantum state. While this may sound almost tautological, what MLE fails to give credit to is a large multitude of states that are almost as likely as $\rho_{\rm MLE}$ (see Ref.~\citep{B-K2010}). Bayesian methods, on the other hand, take these states into account naturally. We will briefly compare MLE and Bayesian methods for entanglement estimation, and in later sections we will get into more details.

Bootstrap methods combined with MLE can be used to generate a distribution of states (somewhat similar to the Bayesian posterior distribution of states). Here one assumes $\rho_{\rm MLE}$ as the real state, from which new sets (of the same size and type as the actual data set) of simulated measurement results are generated. Each such set yields a new MLE state; and thus a distribution of states is generated, which can be used to generate error bars. While this distribution does take into account to some extent the statistical fluctuations, the final estimation of entanglement might still be overly optimistic (see Ref.~\citep{B-K2010} for a clear account).

The fundamental idea behind Bayesian inference follows from Bayes' theorem. Assume $H$ is a hypothesis and $D$ is the observation data. Bayes' rule tells us that the probability for hypothesis $H$ to be true given observation data $D$, also known as the posterior probability, is
\begin{eqnarray}
  P(H|D)=\frac{P(D|H)}{P(D)}P(H).
\end{eqnarray}
$P(H)$ is the \emph{prior probability}, the probability of $H$ prior to the observations of $D$. $P(D|H)$ is the \emph{conditional probability} for $D$ to be observed if H is true; it is also called the  \emph{likelihood} of the hypothesis given the data; and then it is denoted by ${\cal L}(D)$. $P(D)$ is the \emph{marginal probability} for data $D$, which is usually considered as a normalization factor, namely, as the sum of the conditional probabilities over all mutually exclusive hypotheses
\begin{eqnarray}
  P(D)=\sum_j P(D|H_j).
\end{eqnarray}
In our quantum context, the role of hypotheses is played by density matrices $\rho$.

To be more specific now, we will look at a four-qubit system on which we will perform the simulations presented in the paper (generalizations are straightforward). We assume some POVM is measured that can be written  as a tensor product of local measurements, $\{\Pi_k\}$, (because that tends to be the easiest type of measurement to perform in practice). The outcomes of the POVM measurement can then, likewise, be denoted by $\{\Pi_j\otimes\Pi_k\otimes\Pi_m\otimes\Pi_n\}$, and $f_{jkmn}$ is the frequency of getting the outcome $\Pi_j\otimes\Pi_k\otimes\Pi_m\otimes\Pi_n$. The likelihood functional for any state $\rho$ is then by definition
\begin{eqnarray}
  \mathcal{L}(\rho)&=&\prod_{jkmn}\left[{\rm Tr}\left(\rho\Pi_j\otimes\Pi_k\otimes
  \Pi_m\otimes\Pi_n\right)\right]^{Mf_{jkmn}}\nonumber\\
  &=&\prod_{jkmn}(p_{jkmn})^{Mf_{jkmn}},
\end{eqnarray}
where $M$ is the total number of measurements of the POVM, and
\begin{eqnarray}
  p_{jkmn}={\rm Tr}\left(\rho\Pi_j\otimes\Pi_k\otimes\Pi_m\otimes\Pi_n\right)
  \label{p_jkmn}
\end{eqnarray}
is the probability of the outcome
$\Pi_j\otimes\Pi_k\otimes\Pi_m\otimes\Pi_n$, given the state $\rho$.

The (physical) state that saturates the upper bound of $\mathcal{L}$ is called $\rho_{\rm MLE}$. Since the estimation is a single state, which can be considered as a distribution with zero width, it is equivalent to taking the limit $M\rightarrow\infty$. That is, MLE by reporting a single density matrix essentially assumes that the same data would repeat ad infinitum. Bayesian estimation methods yield the same answer as MLE in that limit, and the influence of the prior is eliminated. In the cases where $\rho_{\rm MLE}$ saturates the upper bound of $\mathcal{L}$ in such a way that $p_{jkmn}\neq{\rm Tr}\left(\rho_{\rm MLE}\Pi_j\otimes\Pi_k\otimes\Pi_m\otimes\Pi_n\right)$  \footnote{In this case there is an Hermitian trace-1 operator $\sigma$ that does satisfy $p_{jkmn}={\rm Tr}\left(\sigma\Pi_j\otimes\Pi_k\otimes\Pi_m\otimes\Pi_n\right)$, but $\sigma$ is not positive definite.}, $\rho_{\rm MLE}$ tends to lie on the boundary of the set of physical states \citep{B-K2010}. That is, the state is of non-maximal rank and some eigenvalues are zero. This usually happens when $M$ is ``small.''

The second step of MLE+bootstrap is to simulate a new dataset by using ${\rm Tr}\left(\rho_{\rm MLE}\Pi_j\otimes\Pi_k\otimes\Pi_m\otimes\Pi_n\right)$ as probabilities of measurement outcomes. Repeating this procedure many times (using the same $\rho_{\rm MLE}$) will produce a distribution of measurement outcomes and inferred quantities, and thus  error bars on those quantities. As the new data are generated by $\rho_{\rm MLE}$, entanglement can be easily overestimated if $\rho_{\rm MLE}$ lies on the boundary of the set of physical states, since generically rank-deficient states are more entangled than full-rank states. On the other hand, if $\rho_{\rm MLE}$ is away from the boundary, then the distribution produced this way is expected to resemble the posterior distribution generated by Bayesian methods.

One interesting question is how fast the gap closes up between the two estimates, MLE and Bayesian, as the number of measurements $M$ increases. In fact, this comparison will serve as a (half) criterion for determining how many measurements is ``sufficient,'' provided we choose some ``standard'' prior distribution to be used in Bayesian entanglement estimation.

\section{Preliminaries}
Before we can tackle the main questions of this paper, we need to make several choices, and we need several  definitions. These are all collected in this Section.

\subsection{PRIORS AND MEASURES}\label{sec_prior}

\begin{figure}[t]
  \begin{center}
    \includegraphics[height=95pt]{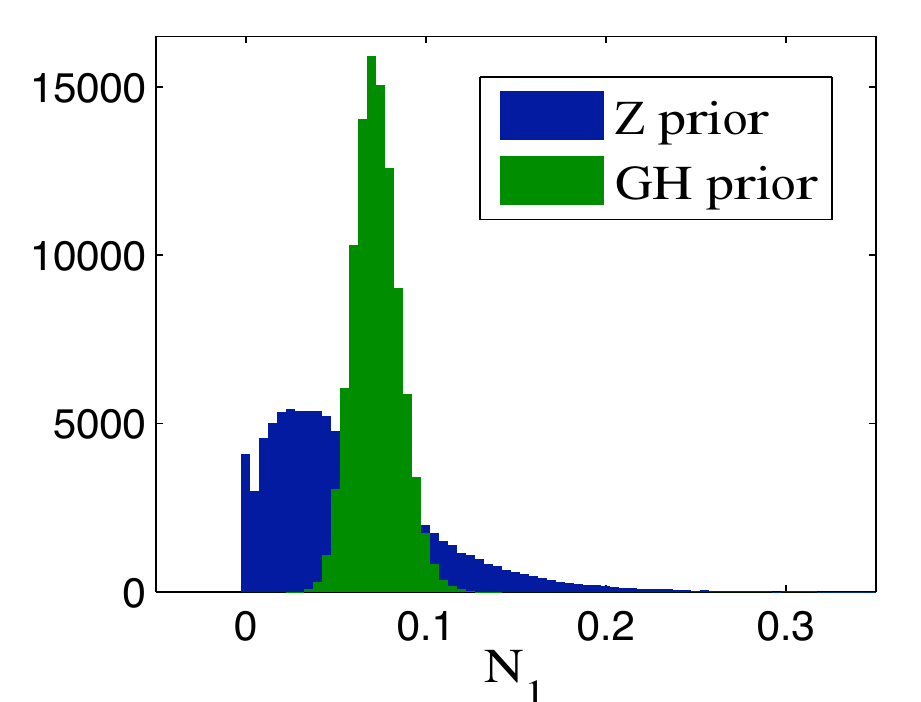}
    \includegraphics[height=95pt]{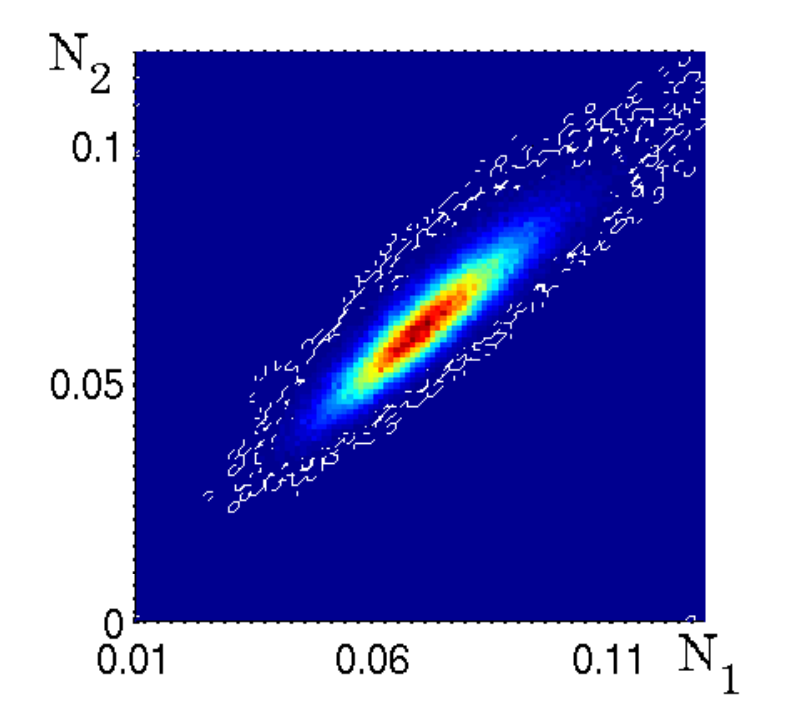}
  \end{center}
  \vspace{-4mm}
  \caption{(Color online) Left: Two different prior distributions of states, named ``Z'' and ``GH'' (for details, see main text, Section \ref{ZGH} in particular), and the induced prior distributions of the 4-party negativity $N_1$ (defined  in Section \ref{sec_multipartite}).
\newline
  Right: GH prior distribution over the two negativities $N_1$ and $N_2$, showing the strong correlation between the two measures for randomly drawn states. The ``white'' noise is due to statistical fluctuations due to the finite sample size.}
  \label{fig_prior}
 \end{figure}

Let $S$ be the set of all physical states $\rho$ and $\mu$ be a measure on the space of $S$. Particularly in probability theory, $\int_S{\rm d}\mu=1$. If $f$ is any real function of $\rho$, then the expectation value of $f$ over the space of $S$ is specified by the measure $\mu$:
\begin{eqnarray}
  \langle f\rangle=\int_S f(\rho){\rm d}\mu.
\end{eqnarray}
If $\rho$ is parameterized by a set of real parameters  $\mathbf{x}$: $\rho=\rho(\mathbf{x})$, then $\mu$ becomes the Lebesgue measure over the space of $\mathbf{x}$: ${\rm d}\mu={\rm d}\mathbf{x}$, where ${\rm d}\mathbf{x}$ is the infinitesimal volume in the corresponding real parameter space. The choice of the parametrization of state $\rho$ , which essentially implies the choice of measure in the space of all states $S$, induces a prior, $P_\mu(\rho)$. Namely, a uniform random distribution over the parameter space defines a particular prior distribution over the space of the physical states through the relation
\begin{equation}
P_\mu(\rho)d\rho=\rho(\mathbf{x})d\mathbf{x}.
\end{equation}
Thus we claim that we have, in this context, established the connection between the prior and the measure.
In numerical implementations where one samples from the random distribution over $\mathbf{x}$ the integral is replaced by the sum:
\begin{eqnarray}
  \int_S f(\rho(\mathbf{x})){\rm d}\mathbf{x}\rightarrow\sum_\mathbf{x}\Delta\mathbf{x}f((\rho(\mathbf{x})).
\end{eqnarray}

\subsubsection{The GH and Z priors}\label{ZGH}

To study a system consisting of four qubits  we choose two inherently different priors: $Z$ and $GH$, which correspond to two distinct measures of the state space. The measures are chosen for their numerical convenience and for their extendability to arbitrary numbers of qubits. Moreover, they are both dense in the set of all states.

To define the $Z$ measure (or prior) we first write the density matrix for a four-qubit system as
\begin{eqnarray}
  \rho=VEV^\dagger,
\end{eqnarray}
where $E$ is a diagonal matrix that carries all the eigenvalues and $V$ is a unitary matrix. The measure of states can be chosen as a product of two particular independent measures introduced in \citep{ZHSL1998,Z1999}
\begin{eqnarray}
  \mu(\rho)=\mu(E)\times\mu(V).
\end{eqnarray}
$\mu(E)$ constitutes a 15-dimensional simplex, which is a uniformly distributed manifold defined by a unit sum of 16 nonnegative numbers, and $\mu(V)$ is the Haar measure based on the direct products of four matrices, any single one of which is to be chosen from the set of three Pauli matrices and the identity. We name the prior corresponding to this measure ``$Z$ prior''.

Alternatively, we can parametrize a four-qubit state as
\begin{eqnarray}
  \rho=HH^\dagger/{\rm Tr}\left(HH^\dagger\right),
\end{eqnarray}
where $H$ is a random complex 16-by-16 matrix, with both the real and the imaginary part of each entry uniformly distributed on $(-1,1)$. This is closely related to Cholesky decomposition of the positive semidefinite matrices \citep{R1992b}, and similar to the parametrization used in Ref.~\citep{BDPS1999}, except that in that paper the unit trace condition is imposed by Lagrangian multipliers while here the condition is satisfied automatically. We name the prior correspond to this measure ``$GH$ prior''. 

Each prior over states induces a prior over any quantity that can be calculated as a function of the state.
If we are interested in a quantity $N(\rho)$, then we have a prior $P(N)dN=(P(\rho) dN/d\rho)d\rho$. In particular, in the next subsection we will define two measures of four-qubit entanglement, two ``negativities'', $N_1$ and $N_2$, that both can be calculated (easily) for given states.
In FIG.\ref{fig_prior} we show the two induced prior distributions over $N_1$ (left) and the $GH$ prior distributions for $N_1$ and $N_2$ (right). From this point on we will stick to $Z$ and $GH$ priors for the demonstration of further results.
Note that these priors are not meant to represent anyone's subjective prior beliefs: rather they are two {\em standard} priors to be used for our specific purposes of quantifying entanglement and determining how many measurement are needed for that.

\begin{figure}[t]
  \begin{center}
    \includegraphics[width=200pt]{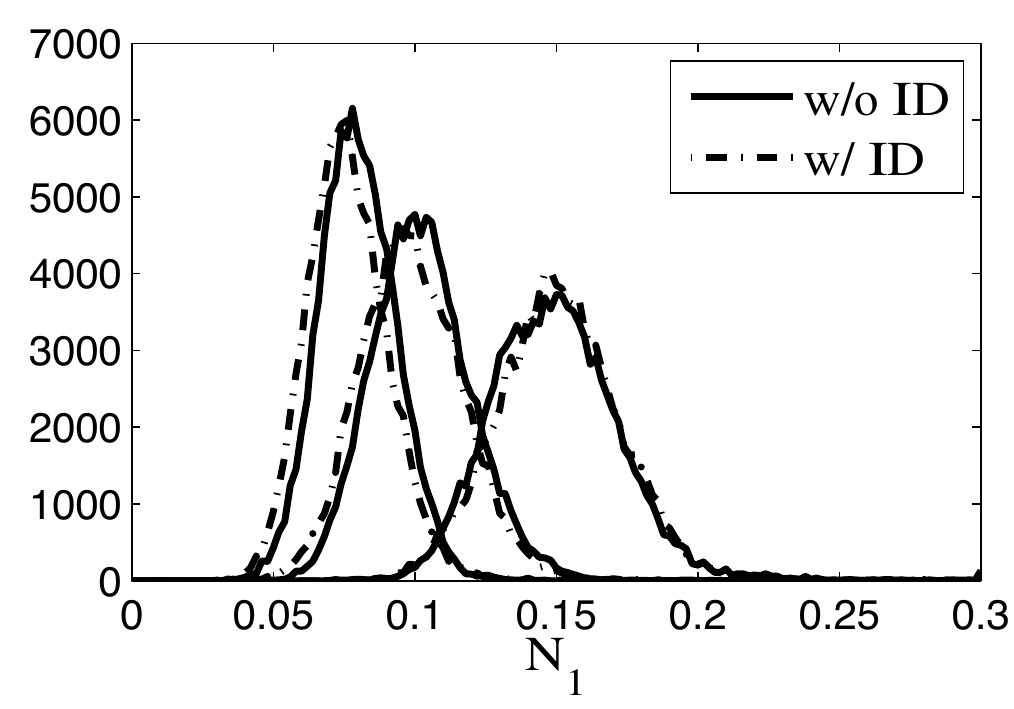}
  \end{center}
  \vspace{-6mm}
  \caption{The posterior distributions resulting from a pure $GH$ prior (w/o ID) and a mixed $GH$ prior (w/ ID, i.e., with the identity mixed in, see text for details). From left to right, the curves describe single trials of just a 1000 measurements on the state $\rho_q$ (Eq.(\ref{rho_q})) with $q=0.4, 0.6, 0.8$ from left to right.}
  \label{fig_N1e3_GHwnID}
\end{figure}

It is worth mentioning some observations on simple variations of the above priors. In particular, both priors have the property that the weight of entangled states is larger than that of unentangled states (more precisely, we compare zero negativity states vs. nonzero negativity states, using our definition of multi-partite negativity, see the following section for details). In order to achieve a prior distribution where the ratio of the weight of separable states vs. entangled states is unity, we can mix in an appropriate amount of the identity matrix into the pure $Z$ and pure $GH$ measures. That is, after having picked a random state $\rho'$ from either measure, we take $\rho=\lambda\rho'+(1-\lambda)\openone/D$, with $\openone/D$ the maximally mixed state in the Hilbert space. $\lambda$ can be sampled from any distribution that leads to a equal weight between entangled and non-entangled states. In our calculation we chose $\lambda=u^{\beta_{Z,GH}}$, where $u$ is uniformly (Lebesgue) random on $[0,1]$ and $\beta_{Z,GH}$ is an adjustable distortion parameter chosen to ensure a 50\% probability of entangled or non-entangled states, $\beta_Z=0.66$ and $\beta_{GH}=0.50$.

How different are the pure and mixed priors as far as quantifying entanglement is concerned? FIG.\ref{fig_N1e3_GHwnID} shows the posterior distributions after just a 1000 measurements for three different states, with pure and mixed $GH$ prior respectively (the plots for the mixed and pure $Z$ distributions are very similar). We find that in every case the ``mixed'' curve gives results very close to the corresponding ''pure'' curve, even when the measurements are still far from sufficient for reliable entanglement quantification (as we will see in Section IV). This indicates that the choice of ``pure'' $Z$ or $GH$ measures is at least somewhat robust against certain simple modifications.

\subsection{MULTIPARTITE ENTANGLEMENT MEASURES}\label{sec_multipartite}

As mentioned, the system we are particularly interested includes four qubits, which is computationally affordable but sufficiently complicated as a step towards scalable multipartite systems. Despite the intensive studies in the multipartite entanglement \citep{BPRST2000,EB2001,Shi2002,VDDV2002,VDM2003,D-DBKM2006,LLSS2007,LLHL2007,WY2008,KK2009,HHK2009,SS2009} over the years, almost all attempts at categorizing multipartite entangled states consider first pure states, and the entanglement measures for pure states can then be extended to mixed states through a convex roof extension, but this involves an arduous minimization over all possible decompositions of the mixed states. To illustrate our ideas without getting too involved in any numerical optimizations, we choose to extend an easily calculable and thereby  desirable measure, namely, negativity \citep{VW2002}, to the four qubit system. The negativity originated from the idea of the partial transpose \citep{P1996}. As is well known by now, for $2\times2$ and $2\times3$ systems that negative partial transpose (NPT) is a necessary and sufficient condition for entanglement \citep{HHH1996}. The negativity has been shown to be closely associated with the fidelity of quantum teleportation \citep{BBCJPW1993} and its logarithm bounds the amount of entanglement that can be distilled \citep{HHH2000}.
The major advantage of the negativity is that it is directly computable for both pure states and mixed states regardless of the size of the system, e.g., the number of qubits, as long as the density matrix is given.

Suppose we have a quantum system consisting of multiple subsystems. We can partition the subsystems into two groups, say $X$ and $Y$. The negativity of a state $\rho$ with respect to that partition $X-Y$, is defined as
\begin{eqnarray}
  \mathcal{N}_{X-Y}(\rho)=||\rho^{\Gamma_Y}||_1-1,
\end{eqnarray}
where $\Gamma_Y$ stands for partial transpose with respect to subsystem $Y$ and $||\cdot||_1$ for the trace norm of a matrix. For four-qubit systems there are two ways of partitioning into groups of certain sizes: ``$2-2$'' (partitioning the four qubits into two groups of two qubits) and ``$1-3$'' (partitioning them into one group of three and one single qubit). Correspondingly we define two negativities as the {\em geometric means}:
\begin{eqnarray}
  &&\mathcal{N}_{2-2}=\left(\mathcal{N}_{AB-CD}\mathcal{N}_{AC-BD}\mathcal{N}_{AD-BC}\right)^{1/3},\\
  &&\mathcal{N}_{1-3}=\left(\mathcal{N}_{A-BCD}\mathcal{N}_{B-CDA}\mathcal{N}_{C-DAB}\mathcal{N}_{D-ABC}\right)^{1/4},\nonumber\\
\end{eqnarray}
where
\begin{eqnarray}
  \mathcal{N}_{AB-CD}=||\rho^{\Gamma_{CD}}||_1-1,
\end{eqnarray}
and similar for all others. For simplicity we henceforth denote $\mathcal{N}_{2-2}$ by $N_1$ and we use $N_2$ for $\mathcal{N}_{1-3}$.

Despite the fact that these negativities can be computed regardless of the system, they do not necessarily make a distinction between all different types of four-party entanglement. For instance, both measures may be nonzero for states that are {\em not} genuinely four-party entangled (e.g., a state like $[\rho_{AB}\otimes\rho_{CD}+\rho_A\otimes\rho_{BCD}]/2$, where $\rho_{AB}$, $\rho_{CD}$, and $\rho_{BCD}$ are entangled); and it may be zero for certain entangled states, namely those with the property that for at least one partition the entanglement is {\em bound}.

On the other hand, both $N_1$ and $N_2$ are entanglement monotones since each single $\mathcal{N}_{AB-CD}$ or $\mathcal{N}_{A-BCD}$ is an entanglement monotone, as shown in Ref. \citep{VW2002}. Moreover, a vanishing $N_1$ or $N_2$, or equivalently a positive partial transpose (PPT) indicates nondistillability with respect to the corresponding partition \citep{DCLB2000}.

Whereas for generic states $N_1$ and $N_2$ are correlated to a high degree (FIG.\ref{fig_prior}), an illuminating counter-example (showing the independence of the two measures) is the Smolin state \citep{S2001}, given by
\begin{eqnarray}
  \rho&=&\frac{1}{4}(
  |\Psi^+\rangle_{AB}\langle\Psi^+|\otimes|\Psi^+\rangle_{CD}\langle\Psi^+|\nonumber\\
  &&+|\Psi^-\rangle_{AB}\langle\Psi^-|\otimes|\Psi^-\rangle_{CD}\langle\Psi^-|\nonumber\\
  &&+|\Phi^+\rangle_{AB}\langle\Phi^+|\otimes|\Phi^+\rangle_{CD}\langle\Phi^+|\nonumber\\
  &&+|\Phi^-\rangle_{AB}\langle\Phi^-|\otimes|\Phi^-\rangle_{CD}\langle\Phi^-|
  ),
\end{eqnarray}
where
\begin{eqnarray}
  &&|\Psi^\pm\rangle=\frac{1}{\sqrt{2}}\left(|01\rangle\pm|10\rangle\right),\nonumber\\
  &&|\Phi^\pm\rangle=\frac{1}{\sqrt{2}}\left(|00\rangle\pm|11\rangle\right).
\end{eqnarray}
For the Smolin state, $N_1=0$ (it's separable along any 2-2 cut) and $N_2=0.5$ (it's entangled along any 1-3 cut). More specifically, $\mathcal{N}_{AB-CD}$ $=\mathcal{N}_{AC-BD}$ $=\mathcal{N}_{AD-BC}$ $=0$, $\mathcal{N}_{A-BCD}$ $=\mathcal{N}_{B-CDA}$ $=\mathcal{N}_{C-DAB}$ $=\mathcal{N}_{D-ABC}$ $=0.5$. The evaluations of the entanglement reflect perfectly what is shown in  Ref.~\citep{S2001}, that for the Smolin state, entanglement can be distilled between any one of the four qubits and part of the rest of the three qubits, while  there is no entanglement between any two groups of two qubits.

\subsection{SIC-POVM AND THE INVERTED STATE}\label{sec_sic-povm}

For no particular reason we will assume we measure, on each single qubit, a class of tomographic POVMs that is symmetric informationally-complete (the so-called SIC-POVMs), where any pair of two outcome vectors has exactly the same overlap. A single qubit SIC-POVM is formulated as \citep{RBSC2004}
\begin{eqnarray}
  \Pi_\alpha=\frac{1}{2}|\alpha\rangle\langle\alpha|,\hspace{7mm}\alpha=1,2,3,4.
\end{eqnarray}
They are linearly independent, tomographically complete and satisfy the normalization condition
\begin{eqnarray}
  \sum_{\alpha=1}^4\Pi_\alpha=\openone,
\end{eqnarray}
and the symmetry condition
\begin{eqnarray}
  {\rm Tr}(\Pi_\alpha\Pi_\beta)=
  \left\{\begin{aligned}
    &\tfrac{1}{4} & \alpha=\beta\\
    &\tfrac{1}{12} & \alpha\neq\beta
  \end{aligned}\right..
\end{eqnarray}
The four-qubit POVM measurement we refer to is the tensor product of the SIC-POVM on individual qubits so that only local measurements are performed. We label the nonorthogonal compound basis as $M_{jkmn},\vspace{2mm}j,k,m,n=1,2,3,4$ and $M_{jkmn}=\Pi_j\otimes\Pi_k\otimes\Pi_m\otimes\Pi_n$. The linear independence and the completeness of $M_{jkmn}$'s can be inferred from the same properties of the $\Pi_\alpha$'s for a single qubit system. This makes it possible to expand arbitrary density matrices in terms of the $M_{jkmn}$:
\begin{eqnarray}
  \rho=\sum_{jkmn}q_{jkmn}\Pi_j\otimes\Pi_k\otimes\Pi_m\otimes\Pi_n.
  \label{rho}
\end{eqnarray}
Note that the coefficients $q_{jkmn}$ here can be negative without compromising the positivity of $\rho$. In fact, in order for $\rho$ to be an entangled state, at least one of them must be negative (otherwise, Eq.~(\ref{rho}) gives a separable form). With the help of Eq.~(\ref{p_jkmn}) we are able to tomographically reconstruct the state by setting the probabilities equal to the measurement frequencies $p_{jklm}$ and then expressing the coefficients $q_{jkmn}$'s in terms of the probabilities $p_{jkmn}$'s:

{\small
\begin{widetext}
\begin{eqnarray}
  q_{jkmn}&=&6^4p_{jkmn}-6^3\left(\sum_\alpha p_{\alpha kmn}+\sum_\beta p_{j\beta mn}+\sum_\gamma p_{jk\gamma n}+\sum_\delta p_{jkm\delta}\right)\nonumber\\
  &&+6^2\left(\sum_{\alpha\beta}p_{\alpha\beta mn}+\sum_{\alpha\gamma}p_{\alpha k\gamma n}+\sum_{\alpha\delta}p_{\alpha km\delta}
  +\sum_{\beta\gamma}p_{j\beta\gamma n}+\sum_{\beta\delta}p_{j\beta m\delta}+\sum_{\gamma\delta}p_{jk\gamma\delta}\right)\nonumber\\
  &&-6\left(\sum_{\beta\gamma\delta}p_{j\beta\gamma\delta}+\sum_{\alpha\gamma\delta}p_{\alpha k\gamma\delta}
  +\sum_{\alpha\beta\delta}p_{\alpha\beta m\delta}+\sum_{\alpha\beta\gamma}p_{\alpha\beta\gamma n}\right)+1.
  \label{q_jkmn}
\end{eqnarray}
\end{widetext}
}
In an actual experiment where the readout frequencies $f_{jkmn}$ are considered as $p_{jkmn}$, the state reconstructed by Eq.(\ref{q_jkmn}) with $p_{jkmn}=f_{jkmn}$ is called $\rho_{\rm tomo}$, which is equal to $\rho_{\rm MLE}$ if and only if $\rho_{\rm tomo}$ is physical (see \citep{B-K2010}). For the case where it is not physical, $\rho_{\rm MLE}$ can be approximated by setting the negative eigenvalues of $\rho_{\rm tomo}$ equal to zero, followed by a renormalization of the density matrix.

\begin{figure}[t]
  \begin{center}
    \mbox{
      \includegraphics[height=110pt]{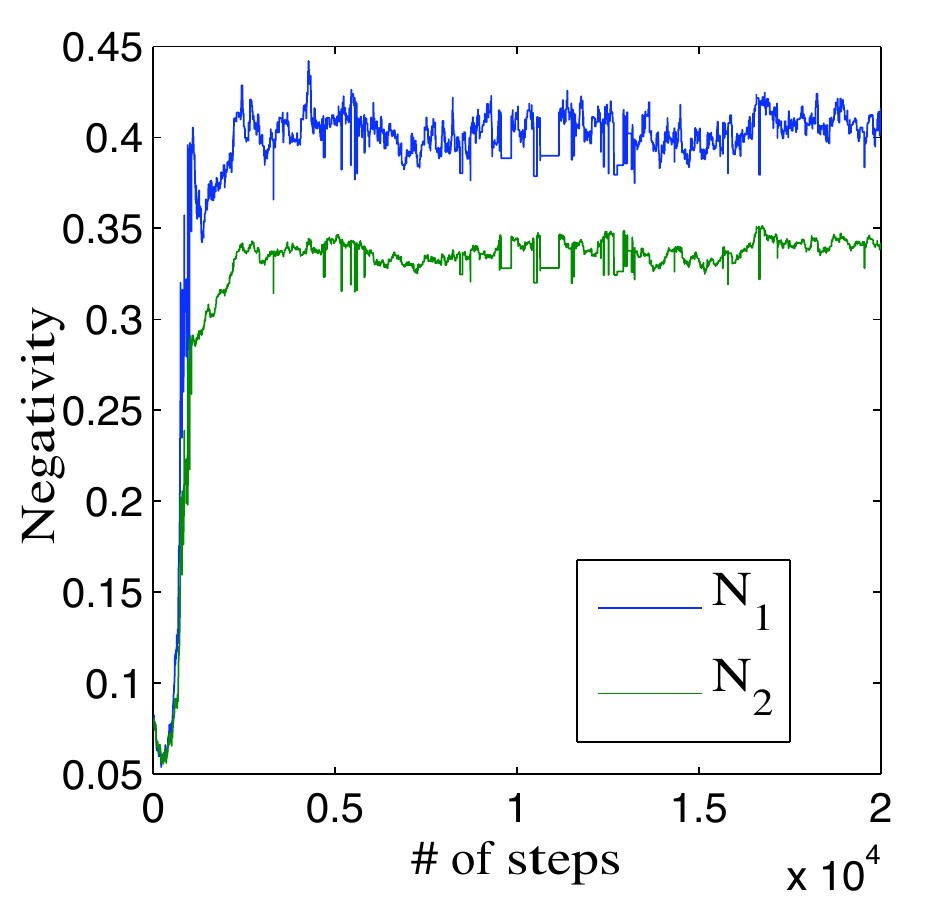}\label{fig_MHwalk1D}
    }
    \mbox{
      \includegraphics[height=110pt]{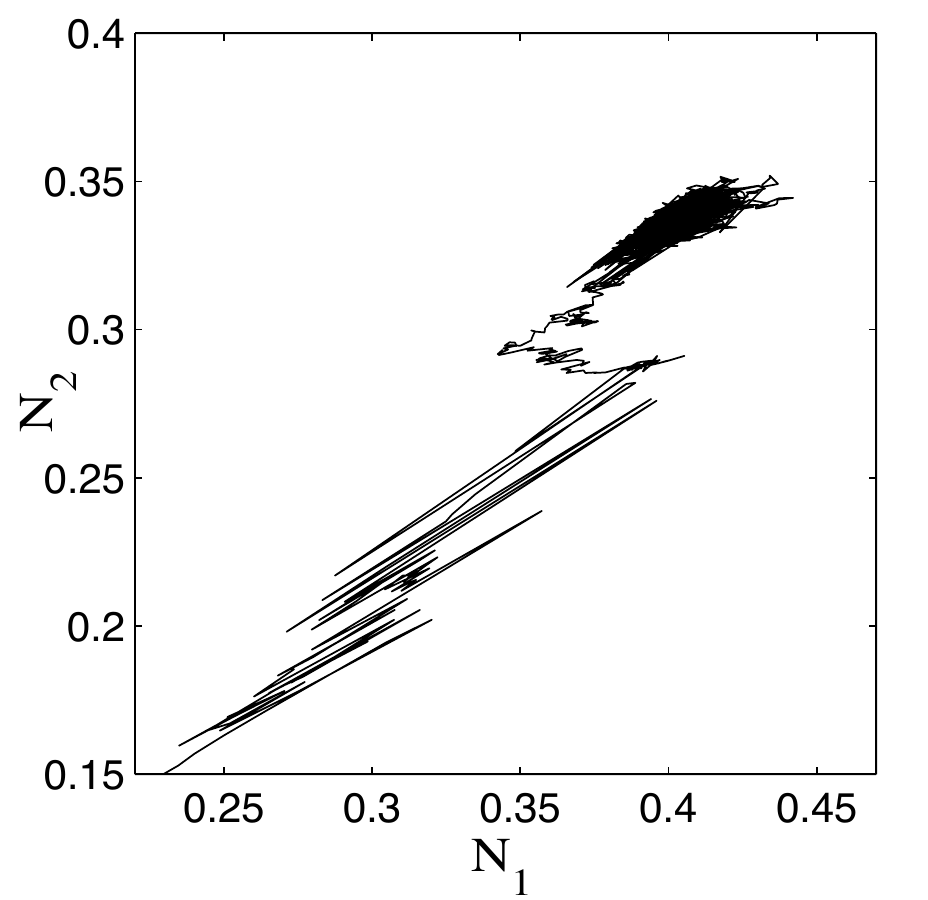}\label{fig_MHwalk2D}
    }
  \end{center}
  \caption{(Color online) A $Z$-prior based Metropolis-Hastings walk towards $\rho(q=0.8)$ (Eq.(\ref{rho_q})), for which $N_1=0.3875$, $N_2=0.3339$. The number of measurements is $10^4$.
   }
  \label{fig_MHwalk}
\end{figure}

\subsection{METROPOLIS-HASTINGS}\label{sec_methods}

The Metropolis-Hastings algorithm (or MH), among many Monte Carlo methods, is applied to generate a Markov chain of states to obtain directly the posterior distribution over states (hence a ``walk'').  The MH walk is known for its fast convergence even when the sampling space is too large for direct random sampling to be efficient. Since the state space of mixed four-qubit states is already  255-dimensional, it makes sense for us to use this method.

 The algorithm starts at any random (physical) state and decides to take (or not take) the following random \emph{step} each time towards a new state depending on the relative likelihood of the new and the old state. The process lasts until a converging distribution is reached from the steps taken. The overall outcome is a path in the state space towards the region with the most likely states and wandering about that region. One then counts how often a certain state occurs; that is its weight in the posterior distribution. More precisely, the probability of taking a step is determined by the ratio of the likelihood of the next and the current state. For example, if the likelihood of the next state is 0.7 times the likelihood of the current state, then there is a chance of 70\% the next state is accepted. On the other hand, if the next state more likely than the current state, i.e. the ratio of the likelihood is larger than 1, then the acceptance is definite. Since the MH walk spends most of its time on the most likely states,  it manages to outperform pure random sampling substantially.

One of the concerns in MH walk is setting the appropriate step size, from one state to the next.  It can be defined in a certain chosen measure as
\begin{eqnarray}
  d_{\rm step}=||\rho_{\rm next}-\rho_{\rm current}||.
\end{eqnarray}
A small step size may costs a long time for the algorithm to converge, although still faster than random sampling, while a large step size tends to identify less likely states by getting stuck in a low likelihood region, which then produces a less accurate distribution.

In standard practice the acceptance rate, which is defined as the overall probability of accepting a step, is used as a quantitative reflection of a step size. There is no rigid proof of what an optimal acceptance rate is, as the final distribution converges to a smooth one. In our work we tested a wide range of possible step sizes, balancing the stability and the efficiency of the program, and managed to keep it between 35\% and 40\%, close to the ideal acceptance rate for Gaussian target distribution \citep{RGG1997}. As shown in FIG.\ref{fig_MHwalk}, the algorithm quickly navigates to the desired area after about 1,000 steps and stays there ``indefinitely'' until we terminate the procedure after $10^5$ steps.

\section{HOW MANY MEASUREMENTS?}\label{sec_measurements}

In order to examine how many measurements suffice for a reliable report of the amount of entanglement in terms of the negativities, it is enlightening to study states that are unlikely to be mistaken as separable states. We choose a particular class of four-qubit states, namely $W$ states with white noise mixed in, which can be characterized by
\begin{eqnarray}
  \rho(q)=q|W\rangle\langle W|+(1-q)\openone/16,\label{rho_q}
\end{eqnarray}
where $|W\rangle=\frac{1}{2}\left(|0001\rangle+|0010\rangle+|0100\rangle+|1000\rangle\right)$ and $\openone$ is the 16-by-16 identity matrix. $|W\rangle$ is known to possess genuine multipartite entanglement \citep{VDDV2002}, and such genuine entanglement can be detected and distinguished from 3-party and 2-party entanglement, as demonstrated recently in an actual experiment \citep{PCDLvEK2009}. According to the entanglement monotones given earlier in the paper, $\rho$ becomes 2-2 separable (i.e., $N_1=0$) when $q<0.1112$  and 1-3-separable (i.e., $N_2=0$)  when $q<0.1262$. When a sufficiently large $q$ value is chosen, the state is less likely to be confused as a separable one. Indeed, the similarities shared between our results for the states $\rho(q=0.8, 0.6, 0.4)$ suggests that the conclusions from these three test states can be validly applied to the class of states with a wide range of $q$ values as long as the state is safely entangled.

\begin{figure}[t]
  \begin{center}
    \includegraphics[width=200pt]{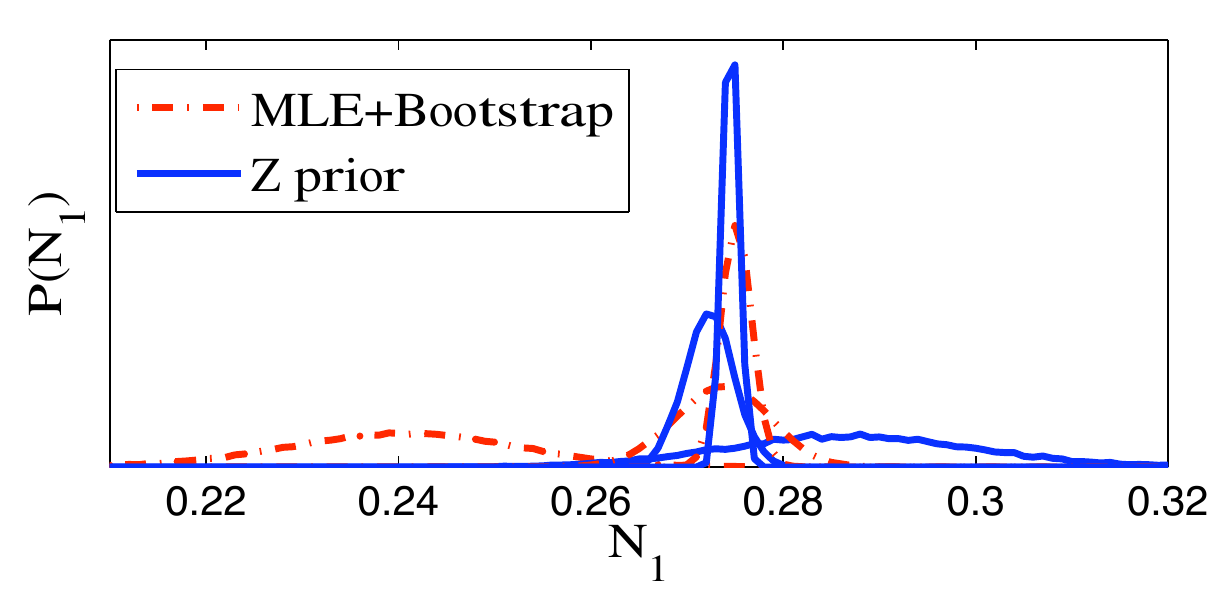}\\
    \includegraphics[width=200pt,]{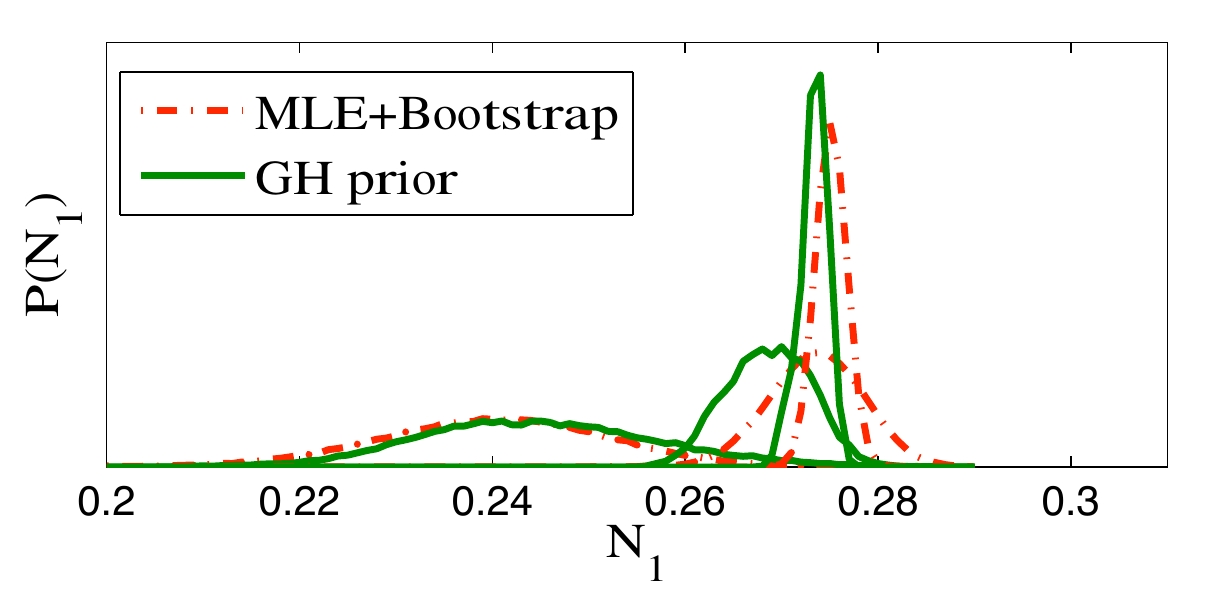}
  \end{center}
  \vspace{-6mm}
  \caption{(Color online) Estimated probability distribution $P(N_1)$ of MLE \& bootstrap method (dot-dashed red) against the posterior distributions with $Z$ and $GH$ priors (solid blue and solid green respectively), after the same series of measurements.  The broadest, the medium and the sharpest distribution for each color correspond to $M=10^4, 10^5, 10^6$. The state being considered is $\rho(q=0.6)$ (Eq.(\ref{rho_q})). The red curves are obtained by assuming $\rho_{\rm MLE}$ is the real state, from which the corresponding measurements are simulated and a $\rho_{\rm MLE}$ is found for each set of measurements, thus not requiring any prior. Around $M\approx 10^5$  measurements all three methods more or less agree with each other.
 }
  \label{fig_postdist06_ZGHMLE}
\end{figure}

\begin{figure}[t]
  \begin{center}
    \includegraphics[height=130pt]{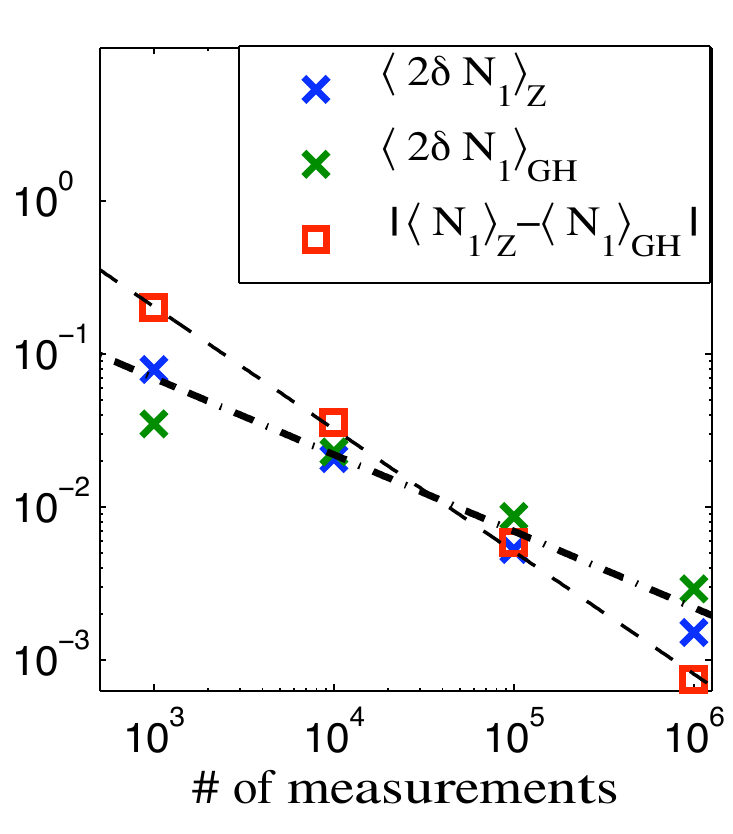}
    \hspace{2mm}
    \includegraphics[height=130pt]{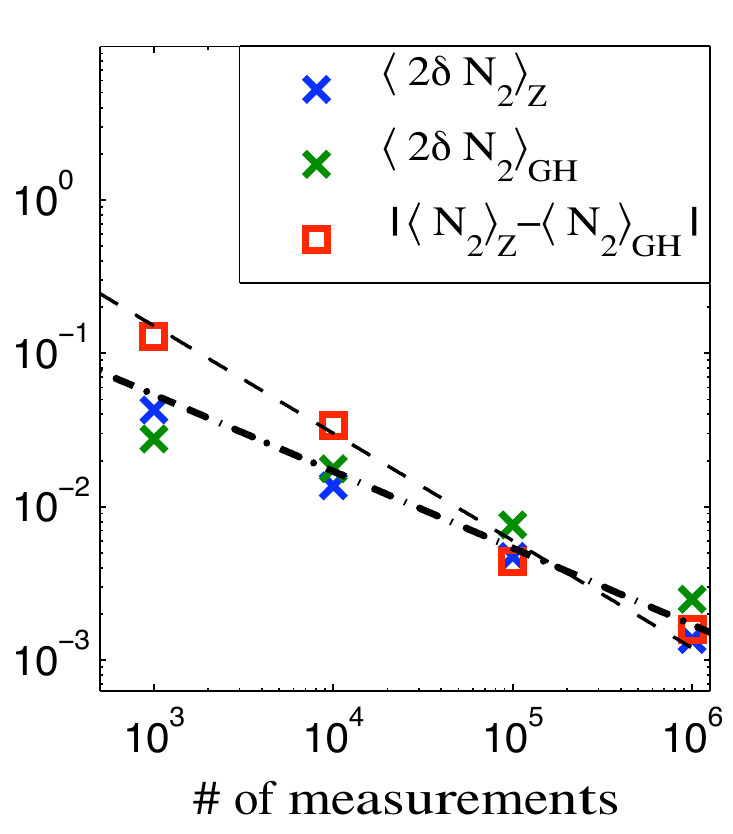}
  \end{center}
  \vspace{-3mm}
  \caption{(Color online) The difference between the estimations of $\langle N_{1,2}\rangle$ using the $Z$ and $GH$ priors, compared with the standard error $\langle\delta N_{1,2}\rangle$ for $\rho(q=0.6)$ (Eq.(\ref{rho_q})). The dash-dotted lines [connecting the sizes of the error bars] are fitted with $c/M^{0.5}$, where $c$ is a number different for the left ($N_1$) and right ($N_2$) figures. The dashed lines [connecting the differences between the two estimates] are fitted to guide the eye with $c/M^{0.81}$ (left) and $c/M^{0.66}$ (right) respectively.
(Figures for $\rho(q=0.8)$ and $\rho(q=0.4)$  look very similar, except for the differences in the slopes of the fitting (dashed) lines.)  Our Criterion 1 is formulated in terms of the average of
$\langle2\delta N_{1}\rangle$
and $\langle2\delta N_{2}\rangle$, such that
the location where the dashed and the dash-dotted lines cross indicates the number of measurements needed for reliable entanglement estimation.   }
  \label{fig_N106stdmeandiffZGHMLE_new}
\end{figure}

In the spirit of Bayesian estimation, the posterior distributions are determined by both the observation data and the prior, with the former becoming more and more important as data accumulates. When the posterior distributions resulting from the two inherently distinct GH and Z priors are laid together, we expect that they will overlap more and more as a function of the number of measurements. Indeed such behavior is demonstrated in FIG.~\ref{fig_postdist06_ZGHMLE} and FIG.~\ref{fig_N106stdmeandiffZGHMLE_new}, and this behavior forms the basis of our Criterion 1. In particular, FIG.~\ref{fig_postdist06_ZGHMLE} shows the evolution of the Bayesian posterior distributions as the number of measurements $M$ increases along $10^4$, $10^5$ to $10^6$. The expectation values $\langle N_{1,2}\rangle$ and the standard errors $\langle2\delta N_{1,2}\rangle$ are computed and shown in FIG.\ref{fig_N106stdmeandiffZGHMLE_new} on a logarithmic scale for the two priors and different numbers of measurements.  Both $\langle2\delta N_{1,2}\rangle_Z$ and $\langle\delta N_{1,2}\rangle_{GH}$ are fitted with $1/M^{0.5}$ (see Appendix).
On the other hand, in the Figure, $|\langle N_{1,2}\rangle_Z-\langle N_{1,2}\rangle_{GH}|$ is fitted with $M^{-\alpha}$, where $\alpha$ is approximately 0.81 for $N_1$ and approximately 0.66 for $N_2$.  The behavior of $|\langle N_{1,2}\rangle_Z-\langle N_{1,2}\rangle_{GH}|$ is analyzed analytically in the $M\rightarrow\infty$ limit in the Appendix, with several important simplifying assumptions made. It shows that for any not-too-pathological prior, the average posterior value of a physical quantity $N$ approaches the true value $N_{\rm r}$  as
\begin{eqnarray}
  |\langle N\rangle-N_{\rm r}|\sim 1/\sqrt{M},
\end{eqnarray}
when $M$ is large. When any two priors are considered with the same observation data, the difference between the average posterior values of $N$ behaves like
\begin{eqnarray}
  |\langle N\rangle_Z-\langle N\rangle_{GH}|\sim1/M,
\end{eqnarray}
which converges faster by a factor of order $\sqrt{M}$. This is because the uncertainty in the data affects each value of $\langle N\rangle$ for each prior in the same linear fashion, and hence this uncertainty is canceled out when the difference is taken. This observation leads directly to our first Criterion.

For entanglement quantification to be reliable we require a number of measurements $M$ such that for $M$ and larger number of measurements, we have
\begin{eqnarray}
{\rm\bf Criterion\,1:}&&\nonumber\\
  |\langle N\rangle_Z-\langle N\rangle_{GH}|&<&\langle\delta N\rangle_Z+\langle\delta N\rangle_{GH}.
  \label{eq_crit_2prior}
\end{eqnarray}
(Obviously, one can always substitute's one favorite measure of entanglement instead of $N$ to create a new criterion. To repeat, our choice of the negativity is for numerical convenience, as well as the fact our measure can be easily generalized to any number of qubits.)
This means the peaks of the two distributions are closer to each other than their mean standard error. When the two priors are well chosen to be sufficiently distinct, the difference likewise is, presumably, sufficiently large to be spotted. As the measurements accumulate, the distribution will converge towards the true value. And when Eq.(\ref{eq_crit_2prior}) is satisfied, we claim that the measurements suffice to be trusted and the posterior distribution from either of the priors qualifies as the final result.

According to the Appendix we can write $|\langle N\rangle_Z-\langle N\rangle_{GH}|=A/M$ in the large $M$ limit, where $A$ is a constant. We also write  $\langle\delta N\rangle_{Z,GH}=B_{Z,GH}/\sqrt{M}$, where $B_{Z,GH}$ are constants, as indicated by the fitting. Then Eq.(\ref{eq_crit_2prior}) is satisfied for $\forall M>A^2/(B_Z+B_{GH})^2$. This is observed when the number of measurements is larger than about $10^5$ (FIG.~\ref{fig_N106stdmeandiffZGHMLE_new}). Therefore $10^5$ is the number of the SIC-POVM measurements necessary, according to Criterion 1, for an honest assessment of the amount of entanglement in terms of negativities in a four-qubit system.

Note that both of the two priors used in this paper are easily generalized to larger number of qubits or other higher dimensional systems. The inherent difference between the two, which is observed in terms of the negativities for two qubits and four qubits (FIG.\ref{fig_prior}), is expected to persist in similar quantities for larger systems. As a result, the proposed criterion can be extended to multi-qubit systems straightforwardly.

Our next criterion compares estimates of entanglement based on MLE with a Bayesian estimate, using a prior $P$ (either $GH$ or $Z$).
For entanglement quantification to be reliable we require a number of measurements $M$ such that for $M$ and larger number of measurements, we have
\begin{eqnarray}
{\rm\bf Criterion\,1.5:}&&\nonumber\\
  |\langle N\rangle_{P}-N_{\rm MLE}|&<&\langle\delta N\rangle_{P}+\delta N_{\rm MLE},
  \label{eq_crit_25prior}
\end{eqnarray}
where $P$ stands for either $Z$ or $GH$. It is, of course, safest (i.e., most conservative)  to employ {\em both} priors, and pick the larger value of $M$ as sufficient. According to the Appendix, $\langle N\rangle _{P}$ and $N_{\rm MLE}$ approach each other at the rate of $1/M$. The argument that $|N_{\rm MLE}-N_{\rm r}|\sim 1/\sqrt{M}$ can be used to imply that $\delta N_{\rm MLE}\sim 1/\sqrt{M}$, since what $\rho_{\rm r}$ is to $\rho_{\rm MLE}$ is exactly what $\rho_{\rm MLE}$ is to all $\rho$'s that constitute the bootstrap distribution. Therefore, similar to Criterion 1, a number of measurements $M$ can always be found for Criterion 1.5 to be satisfied.

In words, the criterion accepts an estimate of entanglement as reliable if the Bayesian estimate, based on some prior $P$, and the MLE estimate (using the bootstrap method) agree with each other. It's only half a criterion, as a Bayesian should see no reason to accept the MLE estimate as judge for his estimate; nor should a frequentist accept the Bayesian estimate with some randomly picked prior for that purpose! It is presumably a good criterion for agnostics (and in that case, not independent of the first Criterion, as MLE will agree with both Bayesian estimates only if the latter agree with each other).

As Fig.(\ref{fig_postdist06_ZGHMLE}) shows, the bootstrap results [for our particular state tested] bear a greater deal of similarities with the $GH$-based posterior distribution than with the $Z$-based posterior. Thus the latter determines the critical value of $M$.  For this particular case, one finds once again that $M\approx 10^5$ is necessary for reliable entanglement quantification. Thus, here both Criteria agree with each other.

\section{Conclusions}
We formulated criteria to determine a sufficient number of measurements for reliable entanglement quantification.
The main criterion uses two different ``standard'' prior distributions over states, used in a Bayesian analysis of the measurement data. Namely, if the two posterior distributions resulting from two different priors agree on the amount of entanglement (within error bars) then we can declare that our results have converged and, therefore, that they are reliable. A second criterion, not quite independent of the first, compares the results from maximum likelihood estimation (MLE), without using any prior, to the two Bayesian results. If MLE agrees with the two Bayesian estimates, then, again, we can declare the results sufficiently reliable. Obviously, in this case the two Bayesian estimates must also agree with each other, and that is why the second criterion is not independent of the first.

We illustrated these criteria by applying them to a particular set of measurements on four qubits [and then both criteria agreed with each other on what constitutes a sufficient number of measurements], but all our results, including the prior distributions, and the measurements considered, and the criteria themselves easily generalize to more (or fewer) qubits.

In order to perform these calculations, we also proposed four-qubit entanglement monotones (based on the  negativity) that can be calculated for arbitrary mixed states. Those monotones, too, generalize easily to different number of qubits.

In fact, the extendability of both entanglement measures and priors to arbitrary numbers of qubits is the principal reason to choose these particular criteria (given these ingredients, the criteria then take a standard form for distinguishing two (peaked) probability distributions).

The next question to be answered is how the sufficient number of measurements scales with the number of qubits. How one can analyze this question when the Hilbert space is so large that even the Metropolis-Hastings algorithm fails to work relaibly, is the subject of a follow-up paper.

\appendix

{\small
\section{Asymptotic behavior of the expectation value of the posterior distribution}

Suppose we are interested in a particular quantity $N(\rho)$, where $\rho$ is a physical state.
Suppose the number of measurements $M$ is large and the posterior for  $N$, $P(N)$, can be approximated by a normal distribution. Then the estimated value of $N$ is where the maximum of $P(N)$ is. We have
\begin{eqnarray}
  &&\left.\frac{{\rm d} P(N)}{{\rm d} N}\right|_{\rho_{\rm max}}=\left.\left(\left.\frac{{\rm d} P(N(\rho))}{{\rm d} \rho}\right/\frac{{\rm d} N(\rho)}{{\rm d} \rho}\right)\right|_{\rho_{\rm max}}=0\nonumber\\
  &\Longrightarrow&\left.\frac{{\rm d}\log P(\rho)}{{\rm d} \rho}\right|_{\rho_{\rm max}}=0,
  \label{eq_app_0derivative}
\end{eqnarray}
provided that ${\rm d} N/{\rm d} \rho$ is analytical in the range of $\rho$.

Recall that
\begin{eqnarray}
  P(\rho)=\frac{P_o(\rho)\mathcal{L(\rho)}}{\int d\rho' P_o(\rho')\mathcal{L(\rho')}},
\end{eqnarray}
where
\begin{eqnarray}
  \mathcal{L}(\rho)=\prod_j p_j(\rho)^{F_j},
\end{eqnarray}
$p_j(\rho)$ is the probability of the $j$'th result to be observed if the tested state is $\rho$ and $F_j$ is the number of times the $j$'th result is actually observed. We can approximate $F_j$'s in terms of
\begin{eqnarray}
  F_j=Mp_j(\rho_{\rm r})+\sqrt{Mp_j(\rho_{\rm r})[1-p_j(\rho_{\rm r})]}X_j,
\end{eqnarray}
where $\rho_{\rm r}$ is the real state and $X_j$ is a normally distributed variable with variance 1. It means
\begin{eqnarray}
  \overline{X_j}=0, \hspace{2mm} \overline{X_j^2}=1, \hspace{2mm} \textrm{for every $j$}.
  \label{eq_app_X_j}
\end{eqnarray}
The bar average $\overline{X_j}$, instead of the bracket average as in $\langle N\rangle$, indicates that the average is not taken over an ensemble of possible states. Instead, a random $X_j$ value is generated each time a measurement record is collected, as $M$ varies. Whether or not this average is to be taken depends on the specific questions and may be cleared up later. Moreover, $X_j$ and $X_k$ are independent of each other except for one constraint:
\begin{eqnarray}
  \sum_j X_j=1.
\end{eqnarray}
We define
\begin{eqnarray}
  Q(\rho)=\prod_j p_j(\rho)^{p_j(\rho_{\rm r})}
\end{eqnarray}
and
\begin{eqnarray}
  C_M(\rho)=\prod_jp_j(\rho)^{\sqrt{p_j(\rho_{\rm r})[1-p_j(\rho_{\rm r})]}X_j}.
\end{eqnarray}
Then the likelihood function becomes
\begin{eqnarray}
  \mathcal{L}(\rho)=Q^M(\rho)C_M^{\sqrt{M}}(\rho).
\end{eqnarray}
Note that the subscript in $C_M$ suggests the subtle dependence on $M$ through variables $X_j$'s. Hence $C_M(\rho)$ indeed corresponds to a single trial correction. However, note that the behavior we study are not limited to a single trial. In fact, as in FIG.\ref{fig_N106stdmeandiffZGHMLE_new}, the red squares that correspond to $|\langle N_{1,2}\rangle_Z-\langle N_{1,2}\rangle_{GH}|$ really come from multiple trials that are affected by different noise profiles $X_j$'s. Eventually the average over multiple trials is to be taken and the statistics of $X_j$'s will be applied.

Since the integral in the denominator is just a constant, the zero derivative condition Eq.(\ref{eq_app_0derivative}) gives
\begin{eqnarray}
  \left.\left(M\frac{{\rm d}\log Q(\rho)}{{\rm d}\rho}+\sqrt{M}\frac{{\rm d}\log C_M(\rho)}{{\rm d}\rho}
  +\frac{{\rm d}\log P_0(\rho)}{{\rm d}\rho}\right)\right|_{\rho_{\rm max}}=0.\nonumber\\
  \label{eq_app_0derivative1}
\end{eqnarray}
Note that at large $M$, whichever $\rho$ that satisfies Eq.(\ref{eq_app_0derivative}) or Eq.(\ref{eq_app_0derivative1}) is going to be very close to the actual state $\rho_{\rm r}$, which is also the maximum of $\log Q(\rho)$, i.e. $\log Q(\rho_{\rm r})=\log Q_{\rm max}$. We expand $\log Q(\rho)$ around $\rho_{\rm r}$ up to $O\left((\rho-\rho_{\rm r})^2\right)$:
\begin{eqnarray}
  \log Q(\rho)&\simeq&\log Q(\rho_{\rm r})-\frac{1}{2}(\rho-\rho_{\rm r})^T\cdot\left.\frac{{\rm d}^2\log Q(\rho)}{{\rm d}\rho^2}\right|_{\rho_{\rm r}}\cdot(\rho-\rho_{\rm r})\nonumber\\
  &=&L_Q-\frac{1}{2}(\rho-\rho_{\rm r})^T\widetilde{\alpha}(\rho-\rho_{\rm r}),
\end{eqnarray}
where $L_Q=\log Q(\rho_{\rm r})$ and $\widetilde{\alpha}=-\left.{\rm d}^2\log Q(\rho)/{\rm d}\rho^2\right|_{\rho_{\rm r}}$. The first derivative term is absent since $\rho_{\rm r}$ is the local maximum and therefore $\widetilde{\alpha}>0$. This implies
\begin{eqnarray}
  \frac{{\rm d}\log Q(\rho)}{{\rm d} \rho}=-(\rho-\rho_{\rm r})^T\widetilde{\alpha}.
\end{eqnarray}
Similar expansion is applied for $\log P_0(\rho)$ and $\log C_M(\rho)$ so that
\begin{eqnarray}
  \frac{{\rm d}\log P_0(\rho)}{{\rm d} \rho}=\beta^T+(\rho-\rho_{\rm r})^T\widetilde{\gamma},
\end{eqnarray}
where $\beta^T=\left.{\rm d}\log P_0(\rho)/{\rm d}\rho\right|_{\rho_{\rm r}}$ and $\widetilde{\gamma}=\left.{\rm d}^2\log P_0(\rho)/{\rm d}\rho^2\right|_{\rho_{\rm r}}$.
\begin{eqnarray}
  \frac{{\rm d}\log C_M(\rho)}{{\rm d}\rho}=\zeta^T+(\rho-\rho_{\rm r})^T\widetilde{\eta},
\end{eqnarray}
where $\zeta^T=\left.{\rm d}\log C_M(\rho)/{\rm d}\rho\right|_{\rho_{\rm r}}$ and $\widetilde{\eta}=\left.{\rm d}^2\log C_M(\rho)/{\rm d}\rho^2\right|_{\rho_{\rm r}}$.

Therefore the zero derivative condition becomes
\begin{eqnarray}
  -M(\rho-\rho_{\rm r})^T\widetilde{\alpha}+\sqrt{M}\left(\zeta^T+(\rho-\rho_{\rm r})^T\widetilde{\eta}\right)&&\nonumber\\
  +\beta^T+(\rho-\rho_{\rm r})^T\widetilde{\gamma}&=&0.
\end{eqnarray}
Solving it for the maximum state:
\begin{eqnarray}
  \rho_{\rm max}&=&\rho_{\rm r}+\delta\rho,\nonumber
\end{eqnarray}
where
\begin{eqnarray}
  \delta\rho=\frac{1}{\sqrt{M}}\widetilde{\alpha}^{-1}\zeta
  +\frac{1}{M}\widetilde{\alpha}^{-1}(\beta+\widetilde{\eta}\widetilde{\alpha}^{-1}\zeta)
  +O\left(\frac{1}{M^{3/2}}\right).
  \label{eq_app_rho_max}
\end{eqnarray}

Now we suppose that $N(\rho_{\rm max})$ is also where the largest probability of $N(\rho)$ is, which is again assumed to be the expectation value, $\langle N\rangle$. Then
\begin{eqnarray}
  &&N(\rho_{\rm max})\simeq N(\rho_{\rm r})+\left.\frac{{\rm d}N}{{\rm d}\rho}\right|_{\rho_{\rm r}}\cdot\delta\rho\nonumber\\
  &=&N_{\rm r }+\frac{1}{\sqrt{M}}\lambda^T\widetilde{\alpha}^{-1}\zeta
  +\frac{1}{M}\lambda^T\widetilde{\alpha}^{-1}(\beta+\widetilde{\eta}\widetilde{\alpha}^{-1}\zeta)
  +O\left(\frac{1}{M^{3/2}}\right),\nonumber\\
  \label{eq_app_N_max}
\end{eqnarray}
where $N_{\rm r}=N(\rho_{\rm r})$ and $\lambda=\left.{\rm d}N(\rho)/{\rm d}\rho\right|_{\rho_{\rm r}}$. Since $\widetilde{\alpha}$,  $\beta$, $\zeta$, $\widetilde{\eta}$ and $\lambda$ are all fixed and presumably nonzero, the behavior of $\langle N\rangle$ as it approaches its true value $N_{\rm r}$ goes
\begin{eqnarray}
  |\langle N\rangle-N_{\rm r}|\sim1/\sqrt{M},
\end{eqnarray}
in large $M$ limit.

Note that the first correction in Eq.(\ref{eq_app_rho_max}), $\widetilde{\alpha}^{-1}\zeta/\sqrt{M}$ is merely influenced by the fluctuation of the data through $\zeta$ and the shape of the likelihood function through $\widetilde{\alpha}$, which is determined by the measurement setup. It implies a consistent behavior with no regard of the choice of the prior distribution. Yet the second correction does. We label $\rho_{\rm max}$ and $\beta$ with subscript $Z$ or $GH$ to differentiate the priors. We have
\begin{eqnarray}
  \rho_{Z\rm max}-\rho_{GH\rm max}=\frac{\widetilde{\alpha}^{-1}}{M}(\beta_Z-\beta_{GH}).
  \label{eq_app_rho_diffZGH}
\end{eqnarray}
From the previous analysis we realise that the $M$-dependence in the difference in the state $\rho$ will carry on to the difference in the negativity $N$. Eq.~(\ref{eq_app_rho_diffZGH}) indicates
\begin{eqnarray}
  |\langle N\rangle_Z-\langle N\rangle_{GH}|\sim1/M.
\end{eqnarray}

\section{Asymptotic behavior of the maximum likelihood estimation}

Since we are concerned with the large $M$ region, we assume that $\rho_{\rm MLE}$ is not on the boundary, so that it  satisfies
\begin{eqnarray}
  \left.\frac{{\rm d}\log\mathcal{L}(\rho)}{{\rm d}\rho}\right|_{\rho_{\rm MLE}}=0.
\end{eqnarray}
Using the same expansion in the vicinity of the real state $\rho_{\rm r}$ as in the last section, we obtain
\begin{eqnarray}
  \rho_{\rm MLE}=\rho_{\rm r}+\frac{1}{\sqrt{M}}\widetilde{\alpha}^{-1}\zeta+\frac{1}{M}\widetilde{\alpha}^{-1}\widetilde{\eta}\widetilde{\alpha}^{-1}\zeta+O\left(\frac{1}{M^{3/2}}\right).
\end{eqnarray}
Compared to Eq.(\ref{eq_app_rho_max}), the only difference in the higher-order terms is that the term containing $\beta$ is missing.
When the negativity $N(\rho)$ is considered, similar conclusions can be reached:
\begin{eqnarray}
  |N(\rho_{\rm MLE})-N_{\rm r}|\sim 1/\sqrt{M},
\end{eqnarray}
and
\begin{eqnarray}
  |N(\rho_{\rm MLE})-\langle N\rangle_{Z,GH}|\sim 1/M.
\end{eqnarray}

\bibliography{finally}

\end{document}